\documentclass[12pt,preprint]{aastex}

\slugcomment{Accepted by Astrophysical Journal}

\shorttitle{The X-ray Background problem}
\shortauthors{Treister \& Urry}

\begin{document}

\title{AGN Unification and the X-ray Background}

\author{Ezequiel Treister\altaffilmark{1,2,3} and C. Megan Urry\altaffilmark{2,4}}

\altaffiltext{1}{Department of Astronomy, Yale University, P.O. Box 208101,New Haven,CT 06520.}
\altaffiltext{2}{Yale Center for Astronomy \& Astrophysics, Yale University,
P.O. Box 208121, New Haven, CT 06520}
\altaffiltext{3}{Departamento de Astronom\'{\i}a, Universidad de Chile, Casilla 36-D, Santiago, Chile.}
\altaffiltext{4}{Department of Physics, Yale University, P.O. Box 208120, New Haven, CT 06520.} 

\email{treister@astro.yale.edu}

\begin{abstract}
Active Galactic Nuclei (AGN) unification, which implies a
large number of obscured AGN, can explain the optical,
infrared and X-ray content of deep multiwavelength
surveys. Here, we show that the same model also successfully
explains the spectral shape and intensity of the X-ray
background. The simplest possible unified model assumes a
constant ratio of obscured to unobscured AGN, independent of
redshift or luminosity. With almost no free parameters, the
predicted X-ray background agrees remarkably well with
observations. 
Both the observed
properties of deep AGN samples and the X-ray background can be
explained by the same model without invoking different
evolution for obscured and unobscured AGN. Recent
observational evidence shows that the ratio of obscured to unobscured
AGN may depend on luminosity. An AGN unification model that allows
for such a dependence also fits the X-ray background spectrum, 
and the predicted numbers counts in the 2-10 keV
X-ray band are in good agreement with observations. We
present predictions for the source counts in hard X-rays
(10-100 keV), which indicate that future missions like the
Black Hole Finder will observe thousands of heavily obscured AGN.
\end{abstract}

\keywords{galaxies: active --- X-rays: diffuse background}

\section{Introduction}

More than forty years after its discovery, the nature of the
X-ray background emission is still debated. About 80\% of
the background at low energies ($\la 4$ keV) has been
resolved into point sources \citep{mushotzky00}. Above 4 keV
the situation is less clear, especially above 10 keV, where
only a small fraction of the X-ray background has been
resolved \citep{worsley04}. Since obscured AGN have a hard
X-ray spectrum --- harder than the typical unobscured AGN
--- they can be used to explain the observed spectrum of the
X-ray background at energies from a few to $\sim 30$ keV \citep{setti89}.

Early attempts to explain the X-ray background emission
using AGN population synthesis models
\citep{madau94,comastri95,gilli99,pompilio00} needed to
assume a changing ratio of obscured to unobscured AGN with
redshift and/or luminosity. However, recent observational
evidence \citep{gilli04} suggests that this ratio is
constant with redshift. More recent calculations
\citep{ueda03,gandhi03} also assumed a constant
obscured-to-unobscured AGN ratio with redshift; however,
\citet{ueda03} assumed that this ratio changes with
luminosity, while \citet{gandhi03} assumed different
luminosity functions for obscured and unobscured AGN. 
\citet{treister05} pointed out that the previously reported
dependence of this ratio with luminosity, more obscured AGN
at lower luminosities, can arise naturally from selection
effects, for samples extending to high redshift.
More recently, \citet{barger05} presented a much larger
and highly complete sample of X-ray selected AGN in which this
dependence is still observed at low redshifts. 

A general problem for these population synthesis models is
the low redshift peak observed in X-ray selected AGN samples
in deep surveys
(e.g. \citealp{hasinger02,szokoly04,barger03a}), at $z\sim
0.7$, whereas the models predict a peak at $z\sim 2$,
similar to optical quasar surveys \citep{boyle00}. Another
problem is the discrepancy between the observed and
predicted ratios of obscured to unobscured AGN. While most
models need a ratio of 4:1 or higher to make up the hard X-ray peak in the
X-ray background spectrum, the observed ratio in
the deep X-ray surveys based on spectroscopic
identifications is closer to $\sim$ 2:1
(e.g. \citealp{barger03a,szokoly04}). Both problems were
addressed by \citet{treister04}, who showed that a simple 
AGN unification model with constant ratio of 3:1 could account for
the multiwavelength observations of hard X-ray selected AGN
sample in the Great Observatories Origins Deep Survey
(GOODS; \citealp{giavalisco04}) fields.

Here we verify that population synthesis using AGN
unification models with co-evolving populations of obscured
and unobscured AGN can reproduce the X-ray background
spectral shape and intensity in the 1--100~keV energy
range. This is true for both the simplest AGN unification
model, in which the obscured-to-unobscured ratio is constant
with luminosity and redshift, and for a modified unification
model in which this ratio changes with luminosity.  We also
predict the AGN number counts in the hard X-ray 10-100 keV
band, which will be observed by future missions like the
``Black Hole Finder'' Einstein Probe. Specifically, EXIST
\citep{grindlay03} will see as many as 20,000 AGN in all-sky
observations, while pointed missions like NuStar
\citep{harrison04} will see several hundred AGN per square
degree, depending on the final spatial resolution ($\sim
1000$ sources deg$^{-2}$ at $f_{10-100}=2\times
10^{-14}$~erg~cm$^{-2}$s$^{-1}$). These hard X-ray missions
will have the sensitivity to detect all the active
supermassive black holes out to high redshifts, and the most
luminous AGN at any redshift, allowing for the determination
of black hole demographics throughout the Universe.

Throughout this paper we assume
$H_o=70$~kms$^{-1}$Mpc$^{-1}$, $\Omega_m=0.3$ and
$\Omega_\Lambda=0.7$.

\section{Population Synthesis Using a Unified AGN Model}

The simplest AGN unification model assumes that dust and gas 
surrounds the central engine in a toroidal geometry, which
dictates the distribution of $N_H$ column density along the line of sight. 
\citet{treister04} developed a family of self-consistent AGN spectra
for this model, as a function of orientation and intrinsic hard X-ray luminosity,
from which they calculated the expected AGN number counts in optical
and X-ray bands. 
Here we sum the X-ray spectra of AGN,
modified appropriately by absorption, 
and compare to the observed X-ray background.
Although the distribution of $N_H$ is actually continuous, 
we define ``obscured" and ``unobscured" AGN as those with
$N_H$ greater than and less than $10^{22}$~cm$^{-2}$, respectively.
Extensive details of the model assumptions are
given by \citet{treister04} and the parameters used are
discussed below.

\subsection{X-ray Spectra}

AGN X-ray emission above 2 keV can be well represented by an
attenuated power law of the form:

\begin{equation}
\frac{dN(E)}{dE}\propto E^{-\Gamma}e^{-\sigma (E)N_H}e^{-E/E_c} ~,
\end{equation}

\noindent
where $N(E)$ is the number of photons with energy $E$;
$\Gamma$ is the power-law photon index; $\sigma(E)$
represents the cross section for photoelectric absorption of
soft X-rays, given by \citet{morrison83} assuming solar
abundance of metals; $N_H$ is the neutral hydrogen column
density along the line of sight; and $E_c$ is the cutoff
energy. In this work we assume typical values for the
intrinsic power law slope, $\Gamma=1.9$ (e.g.,
\citealp{nandra94,nandra97,mainieri02}), and the cutoff
energy, $E_c=300$~keV \citep{matt99}. Later we explore the
effects of changing these parameters (\S 2.3).

At hard X-ray energies ($E\ga 10$keV), the contribution to AGN spectra
from emission reprocessed by the accretion disk is
significant. To account for this emission we used the
Compton reflection models of \citet{magdziarz95} assuming
solar abundance of metals, a 2 times solar iron abundance
and an average inclination angle with respect to the line of
sight of $\cos i=0.45$; we verified that this angle gives a
spectrum very similar to the average spectrum over all
angles. The assumed iron abundance is consistent with
observations of local AGN, where values of 1-2 times solar
are commonly found \citep{nandra94}, and values as high as
$\sim 3\times$ solar have been measured \citep{boller02}. We
note that for energies below the peak of the X-ray
background, $E\la 30$~keV, a power law spectrum with
$\Gamma=1.7$ is very similar to a spectrum with $\Gamma=1.9$
plus a reflection component \citep{ueda03}.

\subsection{AGN Luminosity function and the $N_H$ Function}

For this work we use the hard X-ray (2-10 keV) AGN luminosity function
and evolution of \citet{ueda03}, which is based on a compilation of
ASCA, HEAO-1 and $Chandra$ observations. Recently, a new hard X-ray
luminosity function was presented by \citet{barger05}, based on a much
larger sample; however, because an analytical fit of the luminosity
function is given only for low redshifts ($z<1.2$), we use the
\citet{ueda03} in this work. 
(In any case, high-redshift luminosity functions at $z>1.2$ are not 
that well-determined, so the analytic fit is necessarily uncertain.) 
We note that in the redshift range over
which AGN contribute significantly to the X-ray background, $z\sim
0.5-1$, the luminosity functions of \citet{ueda03} and
\citet{barger05} are very similar, so the choice of luminosity function
should not affect our results for the synthesized X-ray background.

A key new feature of the \citet{ueda03} luminosity function
is the luminosity-dependent density evolution, in which low
luminosity AGN peak at lower redshift than high luminosity
sources. Similar evolution has been seen in optical quasar
surveys (e.g. \citealp{boyle00}); in that case, the observed
redshift distribution actually peaks at high redshift,
$z\sim 2$, because those surveys are sensitive only to high
luminosity AGN. In contrast, deep X-ray selected samples
include low luminosity AGN, and those with small area, such
as the Chandra Deep Fields, actually exclude high luminosity
AGN because of the limited volume sampled.

\begin{figure}[!ht]
\figurenum{1}
\includegraphics[scale=0.8]{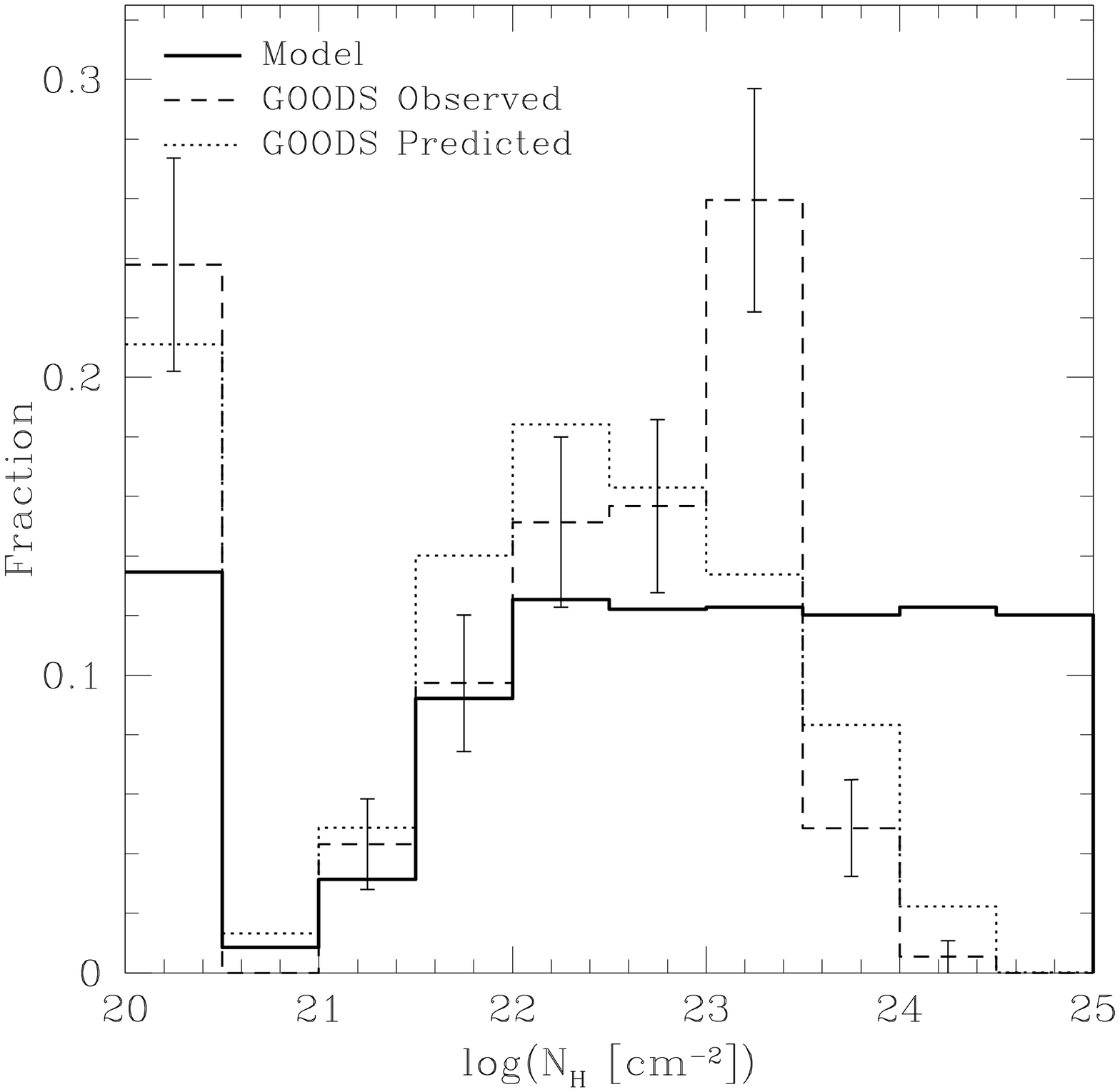}
\caption{Distribution of neutral hydrogen column density ($N_H$) 
along the line of sight calculated assuming the simple
unified model described in the text ({\it thick solid
line}). {\it Dashed line:} Observed $N_H$ distribution for
sources in the GOODS fields as computed by
\citet{treister04} assuming an intrinsic power-law with
$\Gamma=1.7$ and no reflection component. Only Poisson
errors are shown. {\it Dotted line:} $N_H$ distribution
predicted by the model for sources brighter than $2\times
10^{-16}$~erg~cm$^{-2}$s$^{-1}$ in the 2-10 keV band
(roughly the flux limit for the GOODS fields). The agreement
between the observed and predicted distributions is very
good, particularly considering the uncertainties in
determining the $N_H$ value for faint X-ray sources. Highly
obscured sources ($N_H>10^{23.5}$~cm$^{-2}$) are not
detected in the Chandra Deep Fields.
An apparent excess of observed sources at $N_H\sim10^{20}$~cm$^{-2}$
is due to complex X-ray spectra not modeled in our analysis.
}
\label{nh}
\end{figure}

To calculate the number of AGN as a function of $N_H$ we assume a
simple dust torus in which the geometrical parameters give an obscured
to unobscured AGN ratio of $\sim$ 3:1 \citep{treister04}, consistent
with observations of AGN in the local Universe \citep{risaliti99}. In
order to account for the presence of Compton-thick sources we
increased the equatorial column density of \citet{treister04} from
$N_H=10^{24}$~cm$^{-2}$ to $N_H=10^{25}$~cm$^{-2}$. This does not
affect observations at energies lower than 10 keV so it has no effect
on the results presented by \citet{treister04}. 
The \citet{ueda03} luminosity function is based on Compton-thin AGN, 
and so the normalization applies to AGN with $N_H<10^{24}$~cm${-2}$;
the number of additional Compton-thick AGN 
is then given by the $N_H$ distribution.
This adds $\sim 47\%$ more AGN to the original sample 
of \citet{treister04}, none of
which would be detected in even the deepest $Chandra$
observations. Fig.~\ref{nh} shows that the $N_H$ distribution observed
for the X-ray sources in the GOODS fields \citep{treister04} agrees
well with the predictions of the dusty torus model for the GOODS flux
limit ($f_X>2\times 10^{-16}$~erg~cm$^{-2}$s$^{-1}$), particularly
considering the uncertainties in the $N_H$ determination for the
observed sample and the fact that none of the parameters in the model
were adjusted to obtain a better fit to the observations. 
The only exception is that the number of observed sources 
in the $10^{20}$cm$^{-2}$ $N_H$ bin is significantly higher 
than the predicted value. This is because some sources will 
have soft X-ray excesses, in which case our conversion of 
observed hardness ratio into $N_H$ using a simple power law 
will significantly underestimate the amount
of absorption present in the X-ray spectrum.

\subsection{Free Parameters}

Although in this calculation there are many parameters, the
vast majority are fixed by observations. 
Eight (fixed) parameters describe the 
luminosity function and evolution. 
The ratio of obscured to unobscured AGN is fixed 
by the locally observed value; 
when modified as a function of luminosity, 
two more parameters are added (the luminosity range over
which the percentage of obscured AGN is allowed to vary 
from 0 to 100\%).
The distribution of $N_H$ is constrained so that
the equatorial value is $10^{25}$~cm$^{-2}$,
since Compton-thick AGN are observed locally. 
The remaining parameters describe the X-ray
spectrum, and only two of these were adjusted to fit the
observed X-ray background, namely the universal cutoff
energy, $E_c$, and the iron abundance. The choice of cutoff
energy is relevant for $E\ga 80$~keV, while the iron
abundance affects the emission in the $E\sim10-20$ region,
so the two parameters can be determined independently. We
chose the highest cutoff energy that does not overproduce
the observed X-ray background at $E\simeq 100$~keV, namely,
$E_c=300$~keV, and we chose the minimum iron abundance
needed to produce the absorption edge at $E\simeq 10$~keV
while not overproducing the total emission around that
energy. In both cases, the assumed values are in good
agreement with observations of individual AGN, and so cannot
be adjusted arbitrarily. Roughly speaking, acceptable fits
can be obtained for 200 keV$\la E_c\la 400$~keV and iron
abundances in the range $\sim 1.5-3$ solar.

\section{Discussion}

\subsection{The X-ray Background Fit with the Simple Unified Model}

\begin{figure}[!ht]
\figurenum{2}
\includegraphics[scale=0.8]{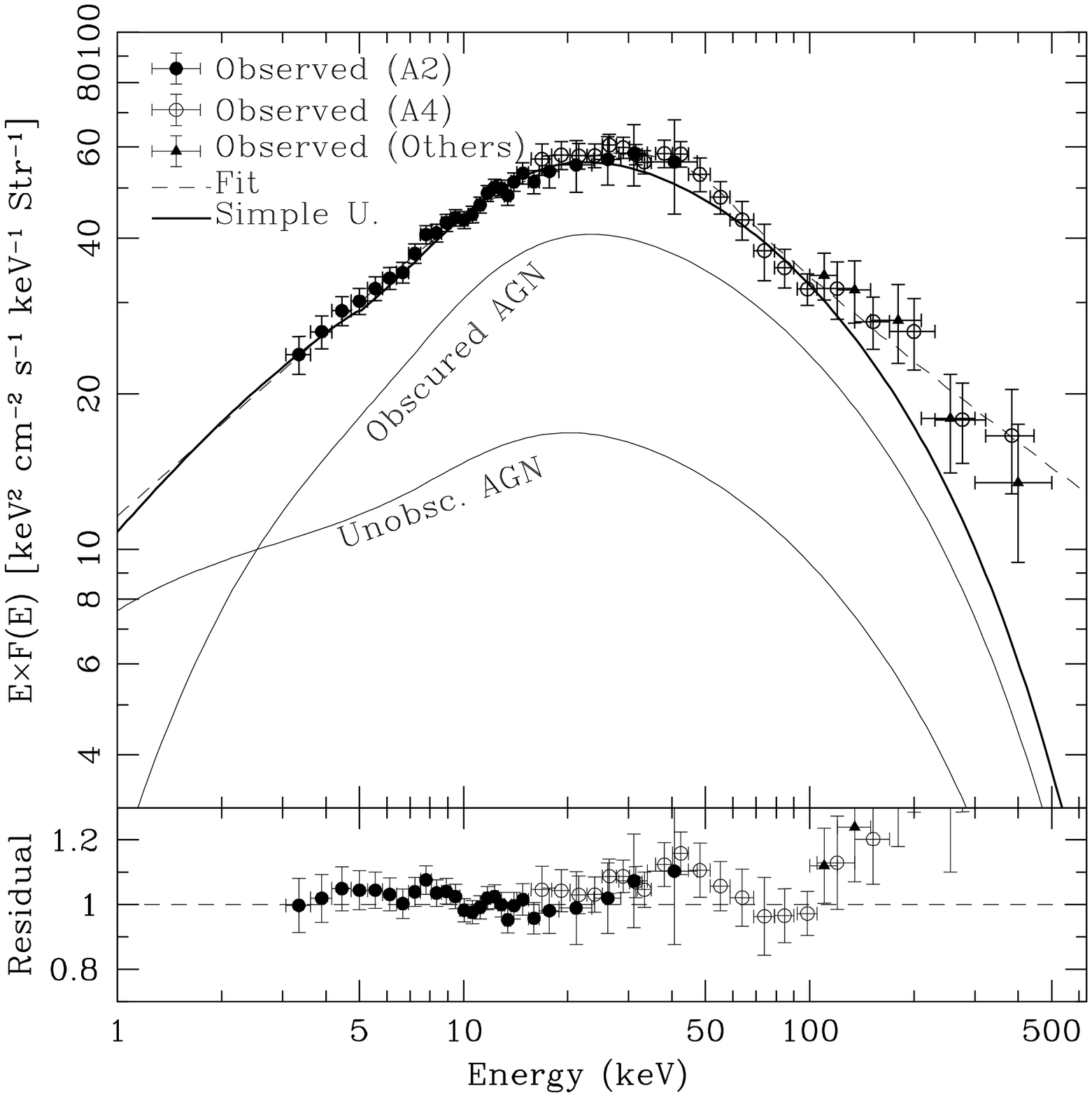}
\caption{{\it Upper panel:} Observed X-ray background 
spectrum ({\it points}) as compiled by \citet{gruber99}
based on HEAO-1 observations. {\it Dashed line:} Analytic
fit to the observed data from \citet{gruber99}. In both
cases the overall normalization was increased by a factor of
1.4 in order to match the observations in the 2-10 keV band,
as described in detail by \citet{deluca04} and \citet{worsley05}.
{\it Thick solid line:} Integrated AGN emission for the simple
unified model described in Section 3.1.
Separate contributions from obscured and unobscured
AGN are shown by the thin solid lines. {\it Bottom panel:}
Residuals (observed/fit). 
the quality of the fit below 100~keV is excellent 
($\chi^2$/DOF=0.794).
At $E>100$~keV, unresolved blazars dominate the X-ray
background (see \citealp{gruber99} and references therein).}
\label{xrb}
\end{figure}

The simplest unified AGN model we consider has a fixed 3:1 ratio of 
obscured to unobscured AGN, independent of redshift or luminosity.
We calculate the total X-ray emission from the AGN population, 
integrating over luminosities
$10^{41.5}$~erg~s$^{-1}<L_X<10^{48}$~erg~s$^{-1}$, 
redshifts $0<z<5$,
and columns densities $10^{20}$~cm$^{-2}<N_H<10^{25}$~cm$^{-2}$. 
Sources with $L_X<10^{41.5}$~erg~s$^{-1}$ or $z>5$ do not contribute
significantly to the X-ray background \citep{ueda03}. For
sources with $N_H>10^{25}$~cm$^{-2}$, the large amount of
obscuration blocks almost all the primary emission and makes
their total contribution to the X-ray background negligible
\citep{gilli99}. The contribution of galaxy clusters in the
soft X-ray region ($E\la 1$~keV) was added following the
prescription of \citet{gilli99}; this component is relevant
only for very low energies, $E\sim 1$~keV, and even then
is only $\sim 10$\% of the total X-ray background
emission. The predicted X-ray background spectrum using this
model is presented in Fig.~\ref{xrb}.

We compare these predictions with HEAO-1 observations of the
X-ray background compiled by \citet{gruber99}. More recent
observations at 2-10~keV using XMM \citep{deluca04} agree in
shape but have 40\% higher normalization than the HEAO-1
data \citep{marshall80}. This higher value of the X-ray background
intensity was observed also in other missions like BeppoSAX
\citep{vecchi99} and ASCA \citep{kushino03}. While a full
reanalysis of the \citet{marshall80} HEAO-1 data is beyond
the scope of this paper, we adopt a value of the
normalization that is 40\% higher than the original HEAO-1
data in order to match the XMM data, a procedure 
followed by other authors previously
(e.g., \citealp{ueda03,worsley05}). Consequences of this
assumption have been discussed by other authors. For example,
\citet{barger03b} concluded that if the higher normalization
is assumed there is very good agreement between the XRB
intensity and the number of resolved sources in deep Chandra observations. 
If instead the XRB intensity at high energies is lower, this
will reduce the number of obscured and/or Compton-thick 
AGN needed to explain the observations. 

In Figure~\ref{xrb} our simple population synthesis model is compared
to the observed data points and an analytic fit from \citet{gruber99}. 
The agreement between predictions and observations is remarkably
good especially given the low number of free parameters 
in the calculation; the $\chi^2$ per degree of freedom for
$E<100$~keV is 0.794.
The residuals are very small below $E>100$~keV; at higher
energies, unresolved blazars (not included here) become dominant (see
\citealp{gruber99} and references therein). 
There is a small discrepancy at $E\sim 40$~keV, 
$\lesssim 5\sigma$, which can be greatly reduced by adding a few ($\sim
5-10\%$) obscured AGN with super-solar iron abundances (which
makes the reflection component peak at higher
energies; \citealp{wilman99}). However, given the low
significance of this discrepancy and the {\it ad hoc} nature
of the solution, we did not add these sources to our model.

This is the first time the X-ray background has been well
fitted by a model fully consistent with the simplest AGN
unification paradigm, i.e., with a constant ratio of
obscured to unobscured AGN, independent of intrinsic
luminosity or redshift. As shown by the thin solid lines on
Fig.~\ref{xrb}, unobscured AGN dominate the energy budget in
the soft X-ray regime ($E\la 3$ keV) while obscured AGN
clearly dominate at higher energies. A similar conclusion
was presented by \citet{bauer04} using a detailed analysis
of the AGN number counts in the Chandra Deep Fields.

\subsection{The Modified Unified Model}

\begin{figure}[!ht]
\figurenum{3}
\includegraphics[scale=0.8]{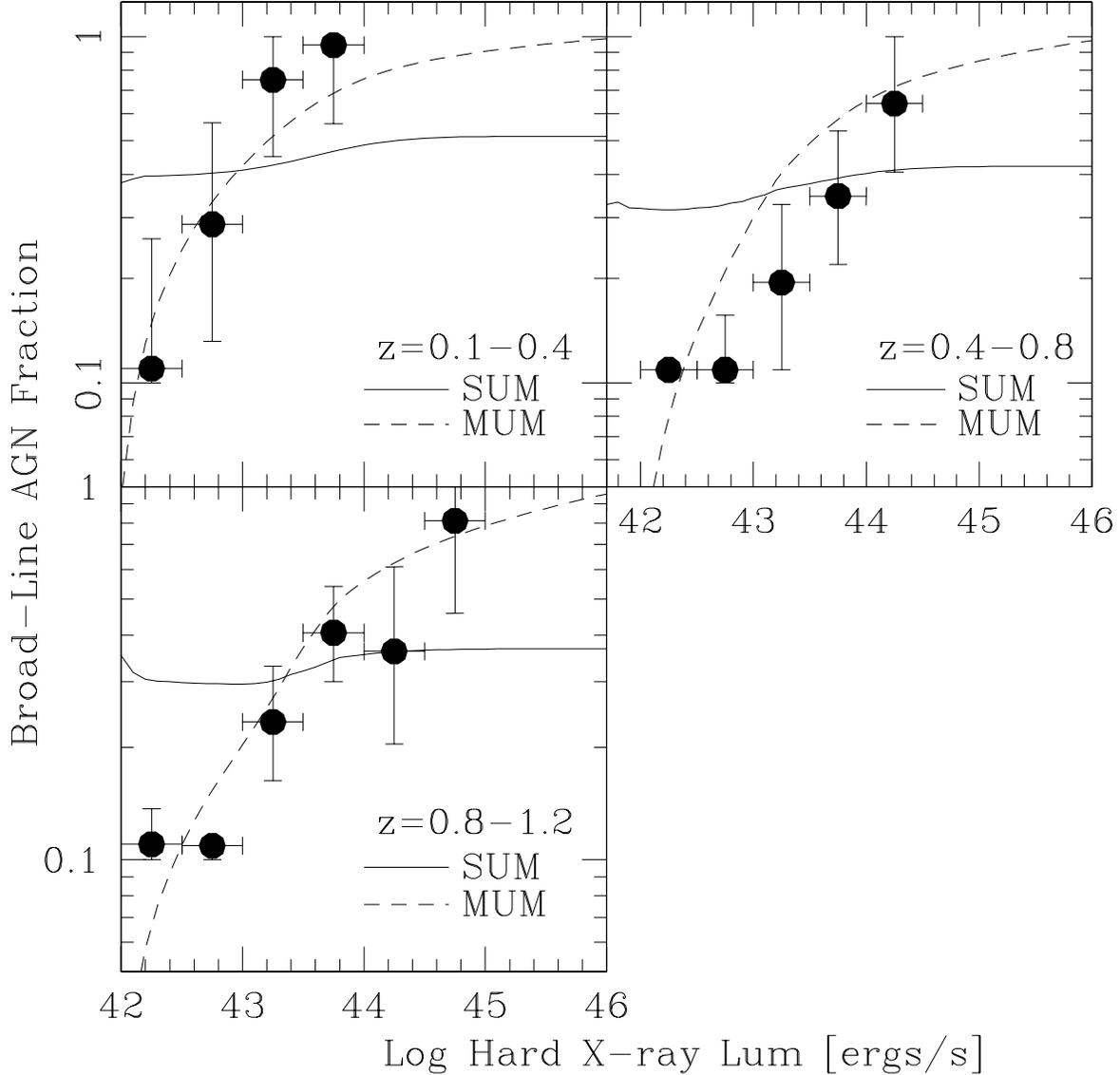}
\caption{Unobscured (broad-line) AGN ratio as a 
function of hard X-rays luminosity in three redshift bins,
z=0.1-0.4 (upper left panel), z=0.4-0.8 (upper right panel)
and z=0.8-1.2 (lower panel). {\it Filled circles}: data
from \citet{barger05}. {\it Solid
lines}: Unobscured AGN ratio as predicted by our simple
unified model. The discrepancy between predictions and
observations is very clear. In this redshift range, 
$z\lesssim 1.2$,
the discrepancy cannot be attributed to selection effects
since the observed sample is mostly complete both for
obscured and unobscured AGN. Therefore, we also used a
modified unified model ({\it dashed lines}) in which the
obscured-to-unobscured AGN ratio decreases with luminosity but
is still independent of redshift, i.e., all AGN have the same
evolution, consistent with simple unification by orientation.
The modified unification model agrees well with observations.  }
\label{ratio}
\end{figure}

An apparent dependence of the obscured to unobscured AGN ratio with
luminosity has been reported \citep{hasinger03,ueda03,steffen03}.
One interpretation is that the torus is evaporated by radiation
\citep{lawrence87}. Indeed, the \citet{ueda03} population synthesis
model for the X-ray background differs from the calculation
above because it assumes that there are relatively more
obscured AGN at lower luminosities. \citet{treister05}
showed that this observed trend with luminosity could be
explained as a selection effect, whereby high-redshift
obscured AGN are too faint for optical spectroscopy and thus
do not have known redshifts or luminosities.  This also can
explain the low peak in the redshift distribution in deep
surveys (e.g., \citealp{hasinger03}).
\citet{barger05} have recently reported this dependence of ratio
on luminosity still pertains at low redshift, where
selection effects are not important \citep{steffen03}. (This
is the first time a sample has been large enough to
determine obscured-to-unobscured ratio as a function of
luminosity at low redshift.)  The simple unified model
predicts that the effects of incompleteness are only
important at $z>1$, therefore at lower redshifts this ratio
is predicted to be constant. In Fig. \ref{ratio}, we compare
the predictions of the simple unification model to the
\citet{barger05} observations, with which they clearly
disagree.

In order to account for this effect, we modify the simple
unification model so that the percentage of obscured AGN
varies linearly from 100\% at $L_X=10^{42}$~erg~s$^{-1}$ to
0\% at $L_X=3\times 10^{46}$~erg~s$^{-1}$.  This new
modified unification model agrees very well with the
\citet{barger05} data , as shown in Fig. \ref{ratio}, and is
also consistent with the observed optical and X-ray number
counts in the GOODS fields.  (The model distributions are
very similar to the simple unification case, Figs.~6 and 7
in \citet{treister04}, and so are not reproduced here.)

\begin{figure}[!ht]
\figurenum{4}
\plottwo{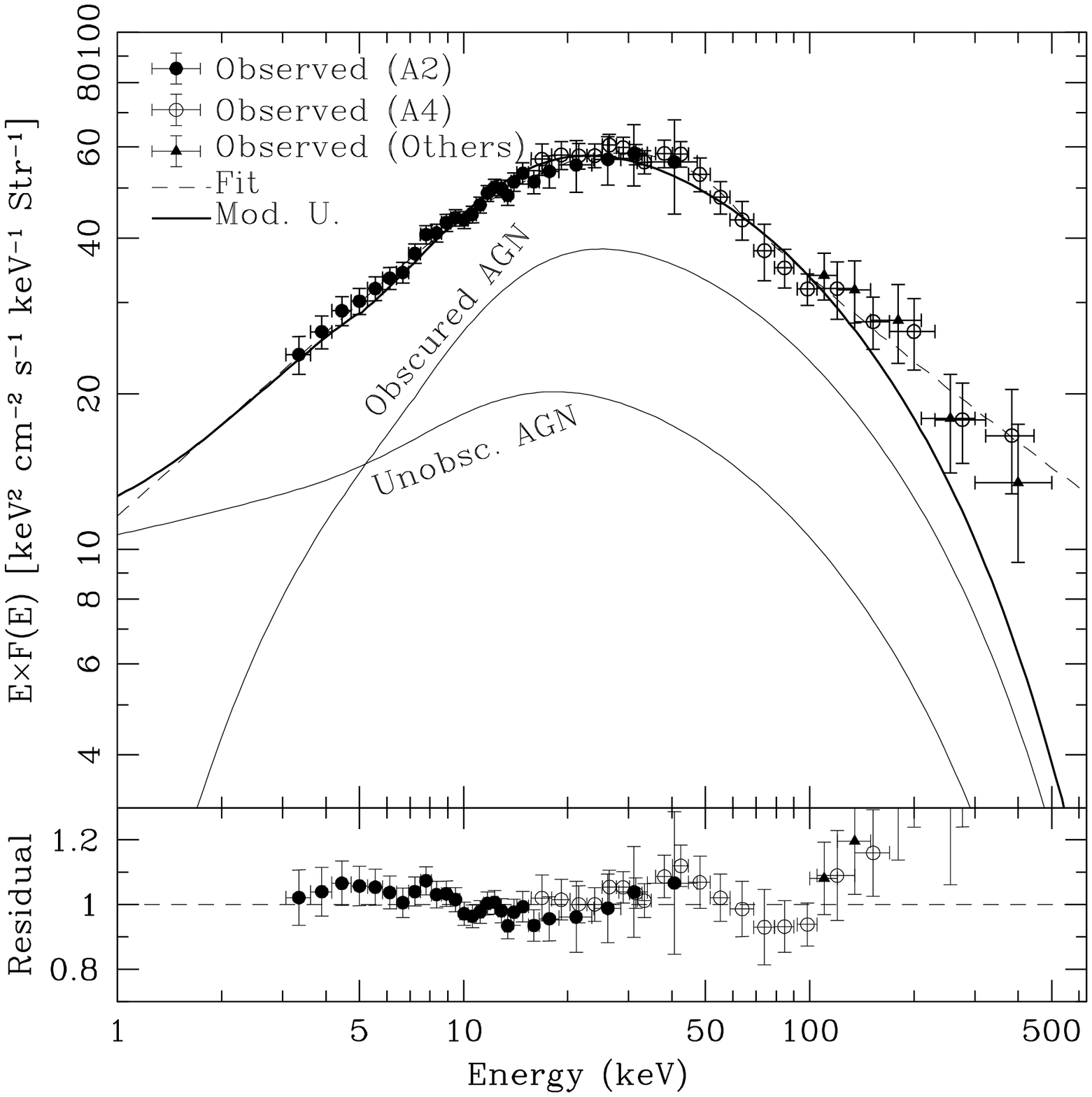}{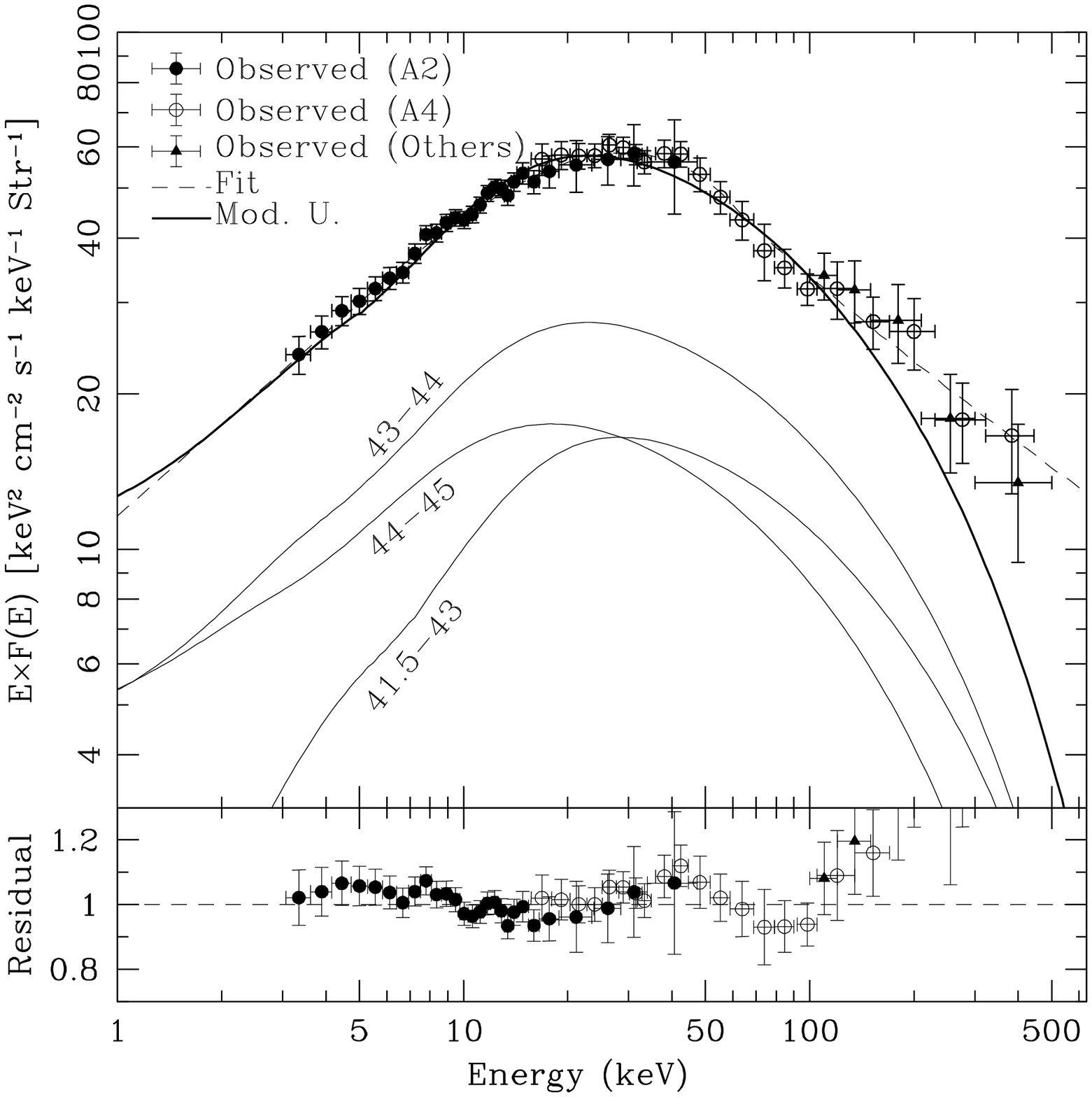}
\caption{{\it Left Panel:} X-ray background population synthesis for the modified 
AGN unification model, with the fraction of obscured AGN
decreasing with increasing luminosity ({\it solid lines}, as
in Fig.~2).  The agreement with observations ({\it data
points, dashed line}, as in Fig.~2) is very good, with a
reduced $\chi^2$ of 0.648. The contribution by unobscured
AGN is more important here than in the simpler unification
model shown in Fig.~2 since obscured AGN are found only at
low luminosities. {\it Right Panel:} Same as left panel, but
showing the contribution from sources in the $\log
L_X$=41.5-43,43-44 and 44-45 bins. The maximum contribution
to the X-ray background comes from sources with $\log
L_X$=43-44, that is, moderate luminosity AGN. }
\label{mod_xrb}
\end{figure}

This modified unified model can also explain the spectral
shape and normalization of the X-ray background, as shown in
Fig. \ref{mod_xrb}; the agreement is good, with a reduced $\chi^2$ of
0.648. According to this model, the contribution of
unobscured AGN is slightly more important than in the simple unified
model since obscured AGN are found only at low luminosities.
The maximum contribution to the X-ray background comes from
moderate luminosity AGN, as shown in Fig.~\ref{mod_xrb}b,
with luminosities in the $10^{43}-10^{44}$~erg~s$^{-1}$
range. A significant contribution also comes from sources
with lower luminosities, while the contribution from
high-luminosity AGN/quasars ($L_X>10^{45}$~erg~s$^{-1}$) is
negligible.



The possible dependence on redshift is less clear. We
combined existing data from the CDF-N, CDF-S and CLASXS
\citep{steffen04} surveys, which have the highest
spectroscopic completeness levels, in order to obtain a
large sample of 189 mostly identified X-ray sources with
$f_X>10^{-14}$~erg~cm$^{-2}$~s$^{-1}$. This sample is
essentially the \citet{barger05} sample minus the 32 bright
ASCA sources, which we excluded in order to have an uniform
coverage down to the X-ray flux limit.

\begin{figure}[!ht]
\figurenum{5}
\plottwo{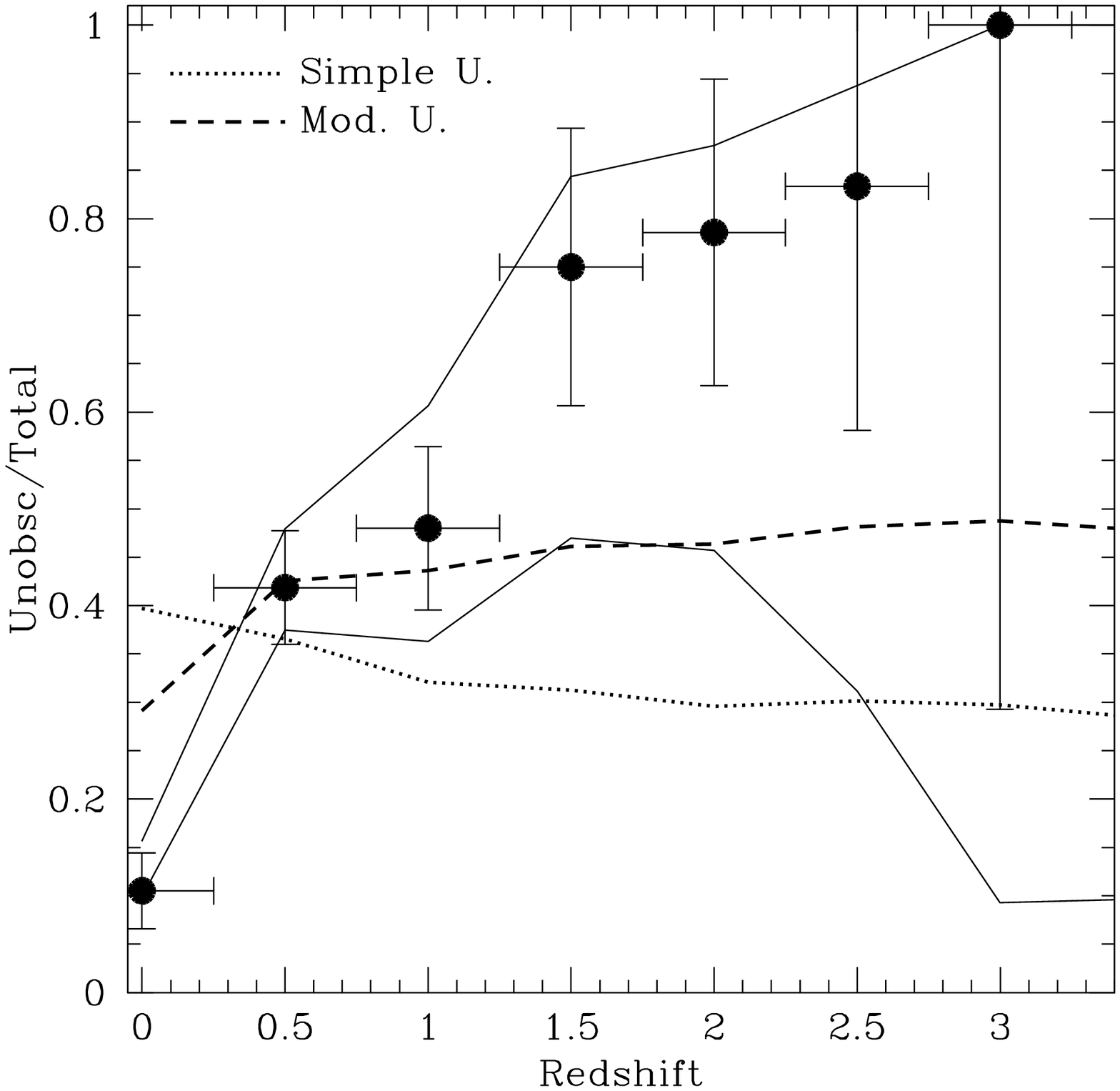}{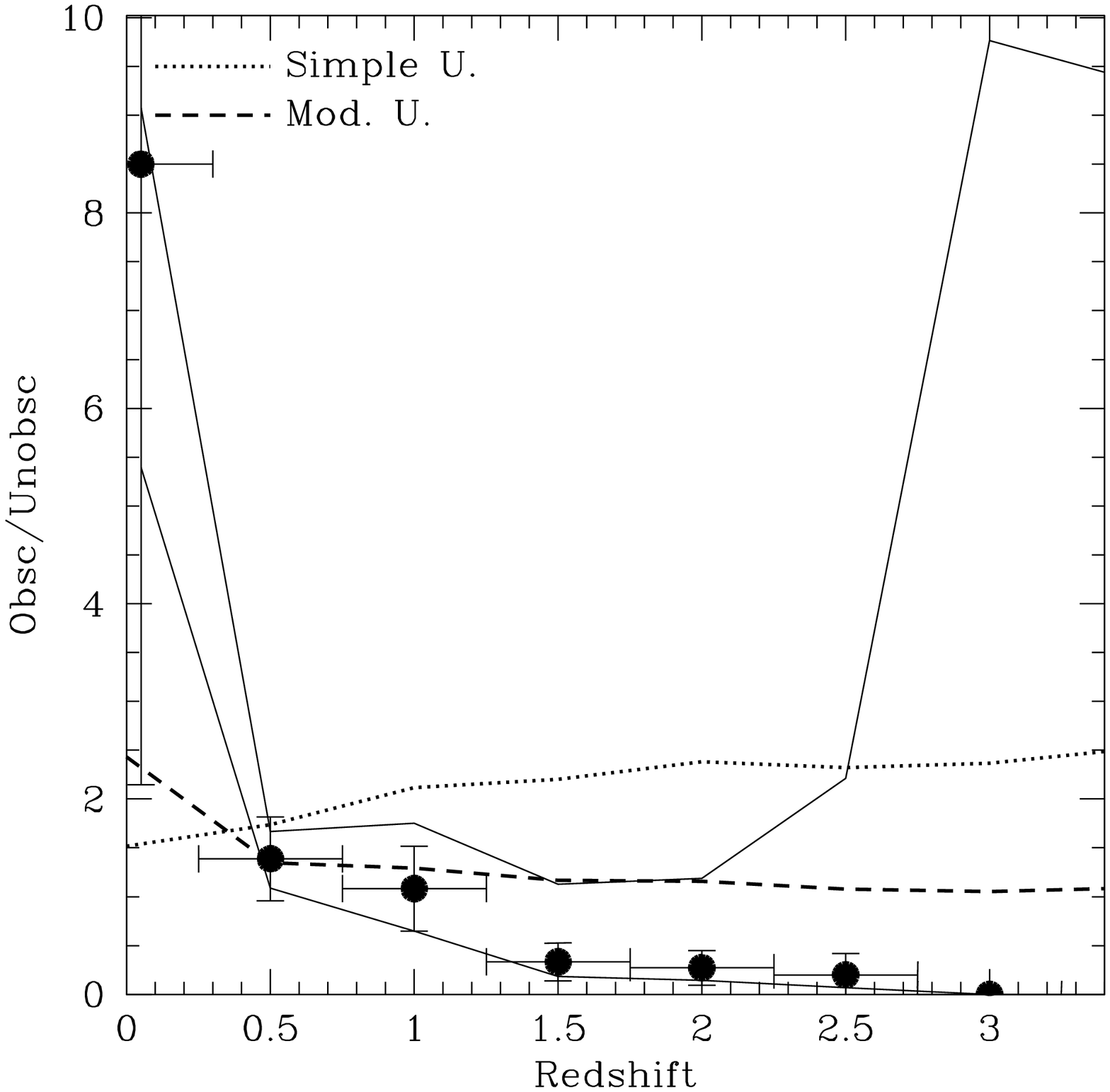}
\caption{
{\it Left Panel:} Ratio of unobscured-to-total AGN as a
function of redshift. The classification is the same as in
Fig.~19 of \citet{barger05}. {\it Circles:} Observed data
points for sources with hard X-ray flux higher than
$10^{-14}$~erg~cm$^{-2}$~s$^{-1}$ for which the spectroscopic
completeness level is 66\%. {\it Thin solid lines:} effects
of adding the unidentified sources to each bin, weighting by
the comoving volume on each redshift bin, assuming that all
the unidentified sources are obscured (lower line) or
unobscured (higher) AGN. {\it Thick lines}: Predicted ratio
as a function of redshift for the simple (dotted) and
modified (dashed) unified models.  {\it Right Panel:} Same
as left panel but for the obscured-to-unobscured AGN ratio
versus redshift.}
\label{rat_red}
\end{figure}

Fig.~\ref{rat_red} shows the observed unobscured-to-total
and obscured-to-unobscured AGN ratios versus redshift, for
sources with hard X-ray flux greater than
$10^{-14}$~erg~cm$^{-2}$~s$^{-1}$. Excluding fainter X-ray
sources means the spectroscopic completeness of this sample
is very high (66\%). The agreement between the modified
model and observations is good up to $z\sim 1$. At higher
redshifts an apparent increase in the unobscured-to-total
ratio with redshift is observed, contrary to the predictions
of both our simple and modified models. However, this may be
due to the difficulty of obtaining spectroscopic redshifts
for obscured AGN at $z>1$ (e.g., \citealp{gandhi03}).

In particular, we considered whether the effects of
incompleteness in this sample could explain the apparent
discrepancy.  The unidentified sources could be either
obscured or unobscured AGN; we looked at the limits on the
data obtained by assigning all the unidentified sources to
either obscured or unobscured sources.  Specifically, for
the ratio-versus-redshift plots, we assigned unidentified
sources (of unknown redshift) to redshift bins by weighting
the volume in that bin relative to the total volume of the
sample.  If they are all unobscured AGN one gets the upper
(lower) line in Figure~\ref{rat_red}a(b); if they are all
obscured AGN one gets the lower (upper) line in
Figure~\ref{rat_red}a(b).
                                                                                        
As Figure~\ref{rat_red} shows, the allowed region is
consistent with the modified unification model, that is,
with a unified scenario in which the obscured-to-unobscured
AGN ratio does not change with redshift.  A similar result
was obtained previously using similar (but less complete)
samples (e.g., see \citealp{gilli04}), with somewhat larger
error bars.  Not until redshifts can be measured for
obscured AGN at high redshift will this issue of a possible
redshift dependence be resolved.  Since spectroscopic
redshifts are impossible at such faint flux levels,
medium-band spectrophotometry is necessary in order to
determine whether obscured and unobscured AGN evolve
differently.  If they do, AGN unification at higher redshift
will be ruled out.

\subsection{AGN Number Counts}

\begin{figure}[!ht]
\figurenum{6}
\includegraphics[scale=0.8]{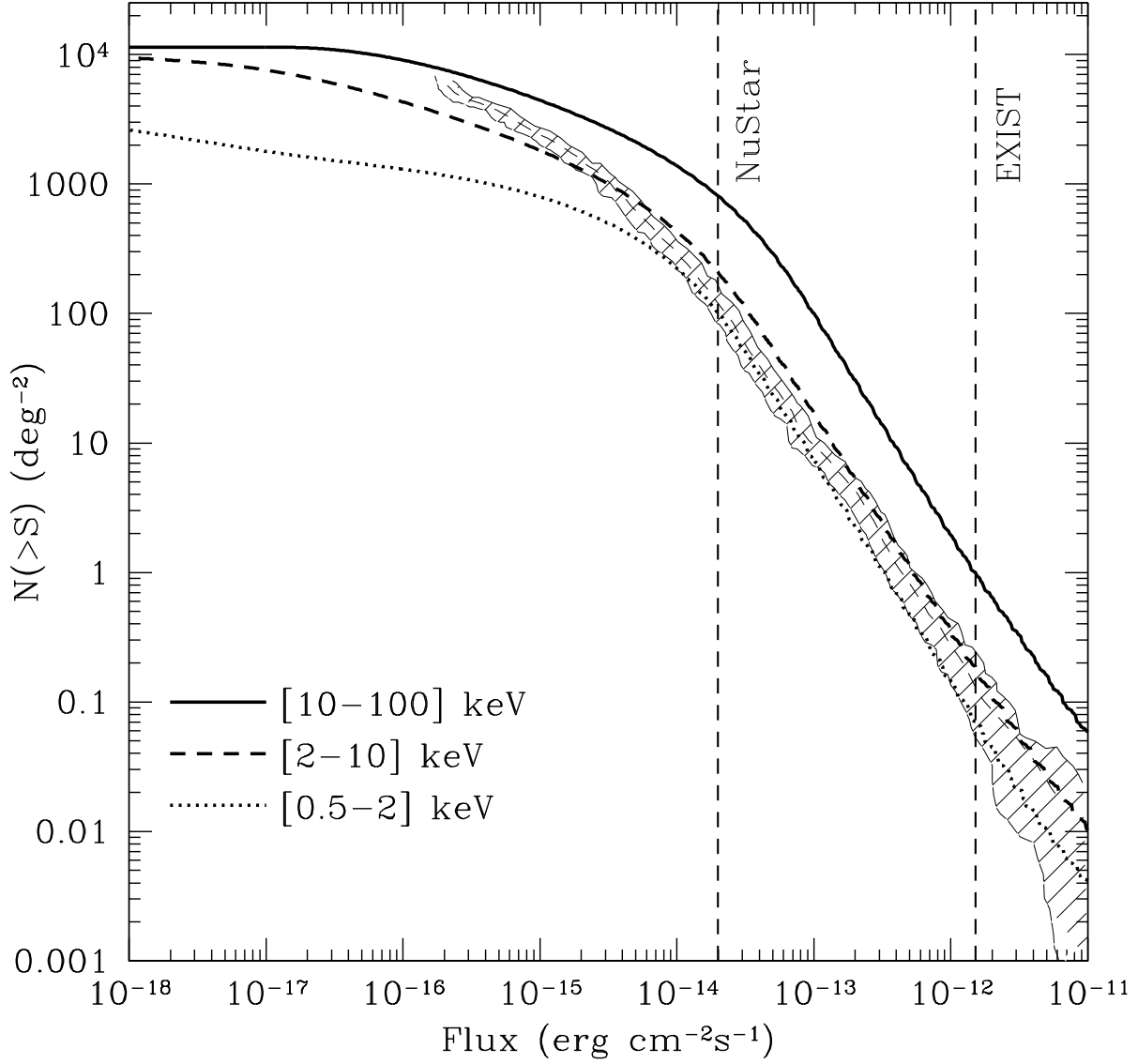}
\caption{Predicted cumulative $\log N-\log S$ distribution
 in the 0.5-2 keV ({\it dotted line}), 2-10 keV ({\it dashed
 line}) and 10-100 keV ({\it solid line}) X-ray bands. {\it
 Shaded region:} Observed $\log N-\log S$ distribution
 (including 1$\sigma$ uncertainties) in the 2-10 keV band
 (\citealp{moretti03}). The agreement between predictions
 and observations is very good. The predicted distribution
 implies that only $\sim 50\%$ of the AGN have been detected
 at the depth of the Chandra Deep Fields. Vertical dashed
 lines show the approximate flux limits in the 10-100 keV
 band for EXIST and NuStar assuming a typical integration
 time of 40~ks.}
\label{ns}
\end{figure}

Figure~\ref{ns} shows the cumulative distribution of sources
as a function of flux (the $\log N-\log S$ plot), calculated
using the modified unification model for three different
X-ray bands: 0.5-2 keV, 2-10 keV and 10-100~keV. (The curves
are not very different for the simple unified model.) The
effects of absorption are most significant in the lowest
energy band, where $\sim 60$\% of the AGN are completely
obscured and thus invisible in soft X-rays. The predicted
number counts in the 2-10 keV bands agree well with the
observed distribution compiled by \citet{moretti03}. At the
flux level of the deepest Chandra observations, this model
suggests that only $\sim 50\%$ of the AGN are detected
\citep{treister04}.

The resolved fraction of the X-ray background in deep
Chandra/XMM surveys goes from $\sim 95$\% in the 0.5-2 keV
band to $\sim 70-80$\% in the 2-7 keV band, to $\la 50$\% at
higher energies \citep{bauer04,worsley04}. The properties of
the unresolved sources are not the same as those of already
detected AGN. Unresolved sources must have a much harder
X-ray spectrum \citep{worsley04} and therefore are likely to
be obscured AGN. These AGN should be detected easily in
surveys at $E\ga 10$~keV, which are unaffected by dust
obscuration, and will be also detected in deep infrared
surveys, where most of the dust re-emission occurs. In
Fig.~\ref{ns} (solid line) we show the predicted $\log
N-\log S$ distribution for 10-100 keV surveys, together with
the expected flux limits for example hard X-ray missions
like EXIST and NuStar. 
At the flux limit of a typical NuStar observation we expect
to have $\sim 800$ AGN per deg$^2$, $\sim$60\% of them
obscured AGN, including $\sim 10$\% Compton-thick. For the
EXIST flux limit, the expected source density is $\sim 1$
AGN per deg$^2$, about 50\% of them obscured AGN ($\sim 6$\%
Compton-thick). EXIST covers a mcuh larger area, so will
detect a much larger sample of AGN, roughly 40,000 all-sky.

\section{Conclusions}

Using the simplest AGN unification model we have explained
the spectral shape and intensity of the X-ray
background. This is the first demonstration that a model
assuming a constant ratio of obscured to unobscured AGN,
independent of redshift or luminosity, can simultaneously
explain the observed X-ray background and the
optical and X-ray counts of AGN detected in deep X-ray
surveys \citep{treister04}. At the same time, a model that
incorporates a changing ratio with luminosity, as suggested
by recently available observations, can also successfully
explain the X-ray background properties. 
The integral constraint of the X-ray background clearly does
not provide a sensitive probe of the fraction of the
obscured AGN; other observations, in particular a high-flux
X-ray sample ($\gtrsim 10^{-14}$~erg~cm$^{-2}$~s$^{-1}$)
with a very high spectroscopic completeness level ($>$90\%),
is needed to test whether the ratio depends on redshift,
i.e., whether the evolution of obscured and unobscured AGN
is different.  At present, unification by orientation --- in
which obscured and unobscured AGN have the same evolution
--- is consistent with the data. In order to obtain this
highly complete sample, accurate redshifts for sources with
optical magnitudes $24\la R\la 27$ are needed. This is
impossible with current state-of-the-art 8m-class
telescopes. However, good photometric redshifts using
medium-band filters down to these magnitudes are possible
and will allow to solve the redshift dependence problem.

The $\log N-\log S$ distribution predicted by the modified
AGN unification model is in very good agreement with
existing observations in the 2-10 keV band. The resolved
fraction of the X-ray background is $\la 50\%$ in the 7-10
keV band and decreases with increasing energy. If the
unification model presented here is correct, $\sim50$\% of
AGN are currently missed by deep {\it Chandra} or {\it XMM}
surveys. These are very obscured AGN that will be detected
only by hard X-ray observatories, like the Black Hole Finder
probe, at X-ray energies where the effects of dust
obscuration are negligible. These surveys will detect a
large fraction of the most obscured AGN, providing for the
first time an unbiased census of the black hole activity in
the Universe.

\acknowledgments

ET would like to thank the support of Fundaci\'on Andes,
Centro de Astrof\'{\i}sica FONDAP and the Sigma-Xi
foundation through a Grant in-aid of Research. This work was
supported in part by NASA grant HST-GO-09425.13-A. We thank
Duane Gruber for providing us his observational data in
electronic format and the anonymous referee for a very
detailed and critical review of this paper and for pointing
our recently available observational results.

\newpage

\end{document}